\documentclass[twocolumn,A4]{article}
\usepackage[dvips]{graphics}
\usepackage{setspace}
\usepackage{color}
\definecolor{gold}{rgb}{0.85,0.66,0}
\definecolor{dblue}{rgb}{0,0,0.8}

\begin{document}

\onecolumn

\begin{center}
{\bf{\Large {\textcolor{gold}{Electron transport through a quantum 
interferometer with side-coupled quantum dots: Green's function
approach}}}}\\
~\\
{\textcolor{dblue}{Santanu K. Maiti}}$^{1,2,*}$ \\
~\\
{\em $^1$Theoretical Condensed Matter Physics Division,
Saha Institute of Nuclear Physics, \\
1/AF, Bidhannagar, Kolkata-700 064, India \\
$^2$Department of Physics, Narasinha Dutt College,
129 Belilious Road, Howrah-711 101, India} \\
{\bf Abstract}
\end{center}
We study electron transport through a quantum interferometer with 
side-coupled quantum dots. The interferometer, threaded by a magnetic 
flux $\phi$, is attached symmetrically to two semi-infinite one-dimensional 
metallic electrodes. The calculations are based on the tight-binding model 
and the Green's function method, which numerically compute the 
conductance-energy and current-voltage characteristics. Our results 
predict that under certain conditions this particular geometry exhibits 
anti-resonant states. These states are specific to the interferometric 
nature of the scattering and do not occur in conventional one-dimensional 
scattering problems of potential barriers. Most importantly we show that, 
such a simple geometric model can also be used as a classical XOR gate, 
where the two gate voltages, viz, $V_a$ and $V_b$, are applied, 
respectively, in the two dots those are treated as the two inputs of the 
XOR gate. For $\phi=\phi_0/2$ ($\phi_0=ch/e$, the elementary flux-quantum),
a high output current ($1$) (in the logical sense) appears if one, and only 
one, of the inputs to the gate is high ($1$), while if both inputs are 
low ($0$) or both are high ($1$), a low output current ($0$) appears. 
It clearly demonstrates the XOR gate behavior and this aspect may be 
utilized in designing the electronic logic gate. 

\vskip 1cm
\begin{flushleft}
{\bf PACS No.}: 73.23.-b; 73.63.Rt. \\
~\\
{\bf Keywords}: Quantum interferometer; Conductance; $I$-$V$ characteristic;
XOR gate; Anti-resonant state.
\end{flushleft}

\vskip 3.8in
\noindent
{\bf ~$^*$Corresponding Author}: Santanu K. Maiti

Electronic mail: santanu.maiti@saha.ac.in

\newpage
\twocolumn

\section{Introduction}

Electronic transport in quantum confined systems like quantum rings, 
quantum dots, quantum wires, etc., has become a very active field both 
in the theoretical and experimental research. The present progress in 
nanoscience and technology has enabled us to use such quantum confined 
geometric models in electronic as well as spintronic engineering since 
these simple looking systems are the basic building blocks of designing 
nano devices. The key idea of manufacturing nano devices is based on 
the concept of quantum interference effect which is generally preserved 
throughout the sample of much smaller sizes, while, it disappears for 
larger systems. A mesoscopic normal metal ring is one such promising 
example where electronic motion is confined, and with the help of such 
a ring we can construct a quantum interferometer. In this article 
we will explore the electron transport properties through a quantum 
interferometer with side-coupled quantum dots, and show how such a 
simple geometric model can be used to design an XOR logic gate. To
reveal this phenomenon we make a bridge system where the interferometer
is attached symmetrically to two external electrodes, the so-called 
electrode-interferometer-electrode bridge. The theoretical 
progress of electron transport in a bridge system has been started 
after the pioneering work of Aviram and Ratner~\cite{aviram}. Later, 
many excellent experiments~\cite{holl,kob,ji,yac,reed1,reed2} have been 
done in several bridge systems to justify the actual mechanisms underlying 
the electron transport. Though in literature many theoretical~\cite{orella1,
orella2,nitzan1,nitzan2,bai,muj1,muj2,zhang,walc2,walc3,cui,gefen,kubo,
naka,aharony,koba,konig,san1,san2,san3} as well as experimental 
papers~\cite{holl,kob,ji,yac,reed1,reed2} on electron transport are 
available, yet lot of discrepancies are still present between the theory 
and experiment.
The electronic transport through the interferometer significantly depends 
on the interferometer-to-electrodes interface structure. By changing the 
geometry one can tune the transmission probability of an electron across 
the interferometer. This is solely due to the quantum interference effect 
among the electronic waves traversing through different arms of the 
interferometer. Furthermore, the electron transport through the 
interferometer can be modulated in other way by tuning the magnetic flux, 
the so-called Aharonov-Bohm (AB) flux, that threads the interferometer. 
The AB flux can change the phases of the wave functions propagating 
along the different arms of the interferometer leading to constructive 
or destructive interferences, and accordingly the transmission amplitude 
changes~\cite{baer2,baer3,tagami,walc1,baer1}. Beside these factors, 
interferometer-to-dots coupling is another important issue that controls 
the electron transport in a meaningful way. All these are the key 
factors which regulate the electron transmission in the 
electrode-interferometer-electrode bridge and these effects have to be 
taken into account properly to reveal the transport mechanisms. 

The purposes of this paper are twofold. In the first part we explore the
appearance of unconventional anti-resonant states for our model. These
states are specific to the interferometric nature of the scattering and
do not appear in ordinary scattering problems. While, the second part 
addresses the XOR gate response in this simple geometry. In our model, the 
interferometer, threaded by a magnetic flux $\phi$, is directly coupled
to two quantum dots, and two gate voltages $V_a$ and $V_b$, are
applied, respectively (see Fig.~\ref{ring}) in these dots. These gate
voltages are treated as the two inputs of the XOR gate. Here we adopt 
a simple tight-binding model to describe the system and all the 
calculations are performed numerically. We narrate the XOR gate behavior 
by studying the conductance-energy and current-voltage characteristics as 
functions of the interferometer-to-dots coupling strengths, magnetic 
flux and gate voltages. Our study reveals that for a particular value 
of the magnetic flux, $\phi=\phi_0/2$, a high output current ($1$) (in 
the logical sense) is available if one, and only one, of the inputs to 
the gate is high ($1$), while if both the inputs are low ($0$) or both 
are high ($1$), a low output current ($0$) appears. This phenomenon 
clearly demonstrates the XOR gate behavior which may be utilized in 
manufacturing the electronic logic gate. To the best of our knowledge 
the XOR gate response in such a simple system has not been described 
earlier in the literature.

The scheme of the paper is as follow. Following the introduction 
(Section $1$), in Section $2$, we present the model and the theoretical 
formulations for our calculations. Section $3$ discusses the significant 
results, and finally, we conclude in Section $4$.

\section{Model and the synopsis of the theoretical background}

\begin{figure}[ht]
{\centering \resizebox*{7.5cm}{7cm}{\includegraphics{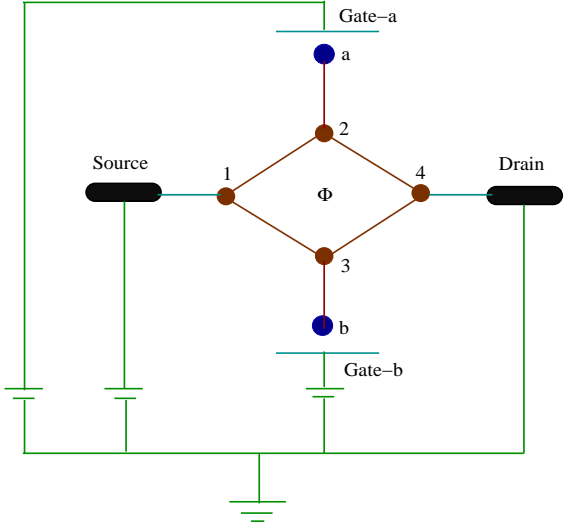}}\par}
\caption{(Color online). The scheme of connections with the batteries for
the operation of the XOR gate. A quantum interferometer with side-coupled
quantum dots (filled blue circles), threaded by a magnetic flux $\phi$,
is attached symmetrically to two semi-infinite $1$D metallic electrodes.
The gate voltages $V_a$ and $V_b$, those are variable, are applied in
the dots $a$ and $b$, respectively. The source and the gate voltages
are applied with respect to the drain. Filled red circles correspond
to the position of the atomic sites in the interferometer.}
\label{ring}
\end{figure}

Let us start by referring to Fig.~\ref{ring}. A quantum interferometer, 
threaded by a magnetic flux $\phi$, is attached symmetrically to two 
semi-infinite one-dimensional ($1$D) metallic electrodes, viz, source 
and drain. Two quantum dots $a$ and $b$ (designed by filled blue circles) 
are directly coupled to the atomic sites $2$ and $3$ of the interferometer, 
respectively. These two quantum dots are subjected to the gate voltages 
$V_a$ and $V_b$, respectively, those are regarded as the two inputs of 
the XOR gate. The gate voltages in the dots are given via the gate
electrodes, viz, gate-a and gate-b. These gate electrodes are ideally
isolated from the dots and can be regarded as two parallel plates of
a capacitor. The actual scheme of connections with the batteries for 
the operation of the XOR gate is clearly presented in the figure 
(Fig.~\ref{ring}), where the source and the gate voltages are applied 
with respect to the drain.

To calculate the conductance of the interferometer with side-coupled 
quantum dots, we use the Landauer conductance formula~\cite{datta,marc}. 
At very low temperature and bias voltage, the conductance $g$ can be 
expressed in terms of the transmission probability $T$ of an electron 
through the interferometer as,
\begin{equation}
g=\frac{2e^2}{h} T
\label{equ1}
\end{equation} 
This transmission probability can be represented in terms of the Green's 
function of the interferometer including the dots and its coupling to 
the two electrodes by the relation~\cite{datta,marc},
\begin{equation}
T={\mbox{Tr}}\left[\Gamma_S G_{I}^r \Gamma_D G_{I}^a\right]
\label{equ2}
\end{equation}
where $G_{I}^r$ and $G_{I}^a$ are respectively the retarded and advanced
Green's functions of the interferometer with the side-attached quantum 
dots including the effects of the two electrodes. The factors $\Gamma_S$ 
and $\Gamma_D$ describe the coupling of the interferometer to the source 
and drain, respectively. For the complete system i.e., the interferometer
with the coupled quantum dots, source and drain, the Green's function 
is defined as,
\begin{equation}
G=\left(\mathcal{E}-H\right)^{-1}
\label{equ3}
\end{equation}
where $\mathcal{E}=E+i\eta$. $E$ is the injecting energy of the source 
electron and $\eta$ gives an infinitesimal imaginary part to $\mathcal{E}$. 
To Evaluate
this Green's function, the inversion of an infinite matrix is needed 
since the full system consists of the interferometer with four atomic 
sites and two coupled quantum dots, and the two semi-infinite $1$D
electrodes. However, the entire system can be partitioned into sub-matrices 
corresponding to the individual sub-systems and the Green's function for 
the interferometer with side-coupled quantum dots can be effectively 
written as,
\begin{equation}
G_I=\left(\mathcal{E}-H_I-\Sigma_S-\Sigma_D\right)^{-1}
\label{equ4}
\end{equation}
where $H_I$ corresponds to the Hamiltonian of the interferometer with 
the two dots. Within the non-interacting picture it can be expressed 
in the form,
\begin{eqnarray}
H_I & = & \sum_i \left(\epsilon_{i} + V_a \delta_{ia} + V_b \delta_{ib} 
\right) c_i^{\dagger} c_i \nonumber \\
 & + & \sum_{<ij>} t \left(c_i^{\dagger} c_j e^{i\theta}+ c_j^{\dagger} 
c_i e^{-i\theta}\right) \nonumber \\
 & + & t_a \left(c_a^{\dagger} c_2 + c_2^{\dagger} c_a \right)
+ t_b \left(c_b^{\dagger} c_3 + c_3^{\dagger} c_b \right)
\label{equ5}
\end{eqnarray}
In this Hamiltonian $\epsilon_{i}$'s are the site energies for all the 
sites $i$ except the sites $i=a$ and $b$ where the gate voltages $V_a$ 
and $V_b$ are applied, those are variable. These gate voltages can be 
incorporated through the site energies as expressed in the above 
Hamiltonian. $c_i^{\dagger}$ ($c_i$) is the creation (annihilation) 
operator of an electron at the site $i$ and $t$ is the hopping integral 
between the nearest-neighbor sites of the interferometer. 
$\theta= \pi \phi/2 \phi_0$ is the phase factor due to the flux $\phi$
threaded by the interferometer. The factors $t_a$ and $t_b$ correspond 
to the coupling strengths of the quantum dots $a$ and $b$ to the atomic 
sites $2$ and $3$ of the interferometer, respectively. For the two
semi-infinite $1$D perfect electrodes, a similar kind of tight-binding 
Hamiltonian is also used, except the phase factor $\theta$, where the 
Hamiltonian is parametrized by constant on-site potential 
$\epsilon^{\prime}$ and nearest-neighbor hopping integral $t^{\prime}$. 
The hopping integral between the source and the interferometer is 
$\tau_S$, while it is $\tau_D$ between the interferometer and the 
drain. The parameters $\Sigma_S$ and $\Sigma_D$ in Eq.~(\ref{equ4}) 
represent the self-energies due to the coupling of the interferometer 
to the source and drain, respectively, where all the information of 
the coupling are included into these two self-energies~\cite{datta}.

The current ($I$) passing through the interferometer is depicted as 
a single-electron scattering process between the two reservoirs of 
charge carriers. The current can be evaluated as a function of the
applied bias voltage by the relation~\cite{datta},
\begin{equation}
I(V)=\frac{e}{\pi \hbar}\int \limits_{E_F-eV/2}^{E_F+eV/2} T(E)~ dE
\label{equ8}
\end{equation}
where $E_F$ is the equilibrium Fermi energy. Here we make a realistic
assumption that the entire voltage is dropped across the 
interferometer-electrode interfaces, and it is examined that under such 
an assumption the $I$-$V$ characteristics do not change their qualitative 
features. 

In this presentation, all the results are computed at absolute zero 
temperature. These results are also valid even for some finite (low)
temperatures, since the broadening of the energy levels of the 
interferometer with side-coupled quantum dots due to its coupling 
with the electrodes becomes much larger than that of the thermal 
broadening~\cite{datta}. On the other hand, at high temperature
limit, all these phenomena completely disappear. This is due to the fact
that the phase coherence length decreases significantly with the rise of
temperature where the contribution comes mainly from the scattering on
phonons, and accordingly, the quantum interference effect vanishes.
Throughout the calculations we set $E_F=0$, and choose the unit $c=e=h=1$.

\section{Results and discussion}

We describe our results in two parts. 
In the first part, we narrate the existence of the anti-resonant states 
in this particular geometry those are specific to the interferometric 
nature of the scattering and do not occur in traditional one-dimensional 
scattering problems of potential barriers. On the other hand, in the 
second part, we demonstrate how this simple geometric model can be used 
as an XOR gate. The key controlling parameter for the whole operation
of the XOR gate is the magnetic flux $\phi$ threaded by the interferometer.

\begin{figure}[ht]
{\centering \resizebox*{8cm}{8cm}{\includegraphics{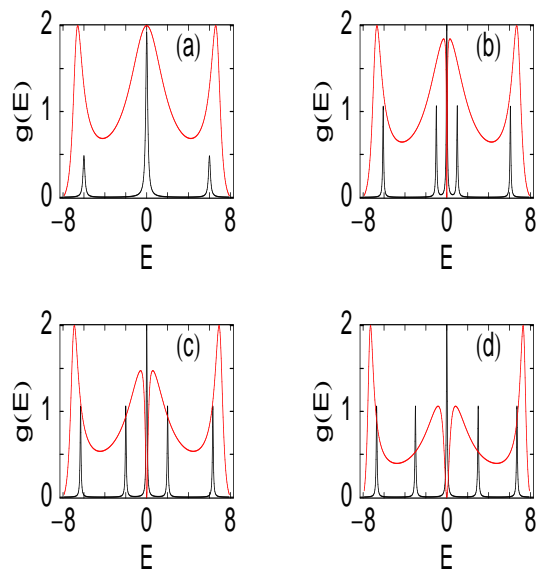}}\par}
\caption{(Color online). $g$-$E$ (red color) and $\rho$-$E$ (black color)
curves for the interferometer with side-coupled quantum dots.
(a) $t_a=t_b=0$, (b) $t_a=t_b=1$, (c) $t_a=t_b=2$ and (d) $t_a=t_b=3$.
Other parameters are, $t=3$, $\tau_S=\tau_D=2.5$, $\phi=0$, and the
on-site potential and the hopping integral in the electrodes are set
as $\epsilon^{\prime}=0$ and $t^{\prime}=4$, respectively. Here we set
$\epsilon_1=\epsilon_2=\epsilon_3=\epsilon_4=0$ and $V_a=V_b=0$.}
\label{cond}
\end{figure}

As illustrative examples, in Fig.~\ref{cond} we present the variation
of the conductance $g$ (red curves) and the density of states $\rho$ (black
curves) as a function of the energy $E$ for the interferometer with 
side-coupled quantum dots, considering the different values of the 
side-coupling strengths $t_a$ and $t_b$, respectively. Here we set the 
site energies of all the atomic sites of the interferometer including the 
two dots as zero. Our results predict that, for some particular energies
the conductances show resonance peaks (red curves), associated with the
density of states (black curves). These energies are the so-called resonant
energies, and the associated states are defined as the resonant states.
The $g$-$E$ spectra predict that, though the resonance peaks are available 
for some particular energies, but the electron conduction from the source
to drain through the interferometer is possible almost for all other 
energies. This is due to the overlap of the two neighboring resonance
peaks, where the contribution for the spreading of the resonance peaks 
comes from the imaginary parts of the self-energies $\Sigma_S$ and 
$\Sigma_D$, respectively~\cite{datta}. At the resonances, the
conductance $g$ approaches the value $2$, and therefore, for these 
energies the transmission probability $T$ goes to unity, since the
\begin{figure}[ht]
{\centering \resizebox*{7cm}{4.2cm}{\includegraphics{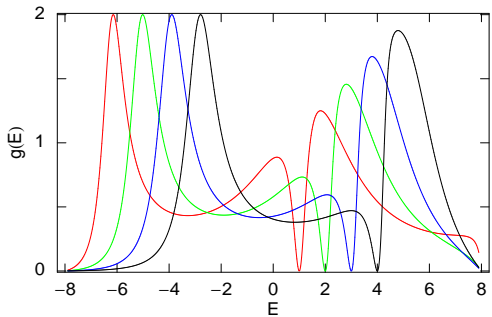}}\par}
\caption{(Color online). $g$-$E$ curves for the interferometer with
side-coupled quantum dots. The red, green, blue and black curves correspond
to the results for the cases when the site energies of all the sites of the
interferometer including the two dots are identically set to $1$, $2$, $3$
and $4$, respectively. Other parameters are, $t=3$, $t_a=t_b=3$,
$\tau_S=\tau_D=2.5$, $\phi=0$, $\epsilon^{\prime}=0$ and $t^{\prime}=4$.}
\label{antipos}
\end{figure}
relation $g=2T$ is satisfied from the Landauer conductance formula 
(see Eq.~(\ref{equ1}) with $e=h=1$). Now we interpret the dependences
of the interferometer-to-dot coupling strengths on the electron transport.
In Fig.~\ref{cond}(a), when there is no coupling between the 
interferometer and the two dots, the conductance shows non-zero value 
for the entire energy range. The situation becomes much more interesting
as long as the coupling of the two dots with the interferometer is
introduced. To illustrate it, in Figs.~\ref{cond}(b)-(d) we plot the
results for the three different choices of the coupling strengths $t_a$
and $t_b$, respectively. The introduction of the side-coupling provides
the anti-resonant state. For all these three different choices of $t_a$
and $t_b$, the conductance spectra (Figs.~\ref{cond}(b)-(d)) show that
the conductance drops exactly to zero at the energy $E=0$. At this
particular energy, the density of states has a sharp peak, and it is 
examined that the height of this particular peak increases very rapidly
as we decrease the imaginary part $\eta$ to the energy $\mathcal{E}$. 
It reveals that the electron conduction 
through this state is no longer possible, and the state is the so-called 
anti-resonant state. With the increase of the side-coupling strength,
the width of the resonance peaks gradually decreases, as clearly observed
from these figures. Both these resonant and anti-resonant states are 
associated with the energy eigenvalues of the interferometer including
the two side-coupled dots, and thus, we can say that the conductance 
spectrum reveals itself the electronic structure of the interferometer 
including the two dots.

To examine the dependence of the anti-resonant state on the site energies 
of the interferometer and the two side-attached quantum dots, in
Fig.~\ref{antipos} we plot the $g$-$E$ characteristics for the four
different cases of these site energies. The red, green, blue and black
curves correspond to the results when the energies of the six atomic
\begin{figure}[ht]
{\centering \resizebox*{8cm}{8cm}{\includegraphics{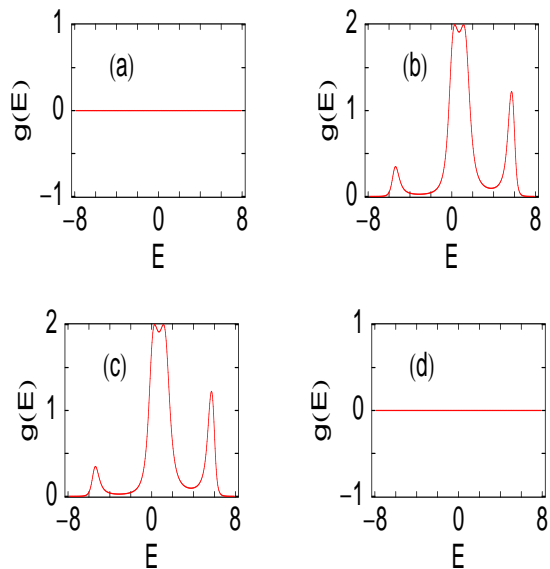}}\par}
\caption{(Color online). $g$-$E$ curves for the interferometer with
side-coupled quantum dots. (a) $V_a=V_b=0$, (b) $V_a=2$ and $V_b=0$,
(c) $V_a=0$ and $V_b=2$ and (d) $V_a=V_b=2$. Other parameters are,
$\epsilon_1=\epsilon_2=\epsilon_3=\epsilon_4=0$, $t=3$, $t_a=t_b=3$,
$\tau_S=\tau_D=2.5$, $\phi=0.5$, $\epsilon^{\prime}=0$ and $t^{\prime}=4$.}
\label{xorcond}
\end{figure}
sites (four sites of the interferometer and the two sites of the dots)
are identically set to $1$, $2$, $3$ and $4$, respectively. Quite 
interestingly we see that the anti-resonant state situates at these
respective energies. Our critical investigation also shows that no
anti-resonant state will appear if the site energy of anyone of these 
six sites is different from the other sites. The similar behavior will
be also observed for the case if anyone of the three hopping strengths 
$t$, $t_a$ and $t_b$ is different from the other two. Thus it can be
emphasized that the anti-resonant state will appear only when the site 
energies of all the six sites are identical to each other as well as 
the strengths of the three different hopping parameters ($t$, $t_a$ and 
$t_b$) are same. In this context it is also important to note that, 
though all the results presented above are done only for the flux 
$\phi=0$, but the positions of these anti-resonant states will not 
change at all in the presence of $\phi$ and no new significant feature 
will be observed for the description of the anti-resonant states. 

Next we concentrate our study on the XOR gate response exhibited by this
geometric model. For the operation of the XOR gate, we set the magnetic 
\begin{figure}[ht]
{\centering \resizebox*{8cm}{8cm}{\includegraphics{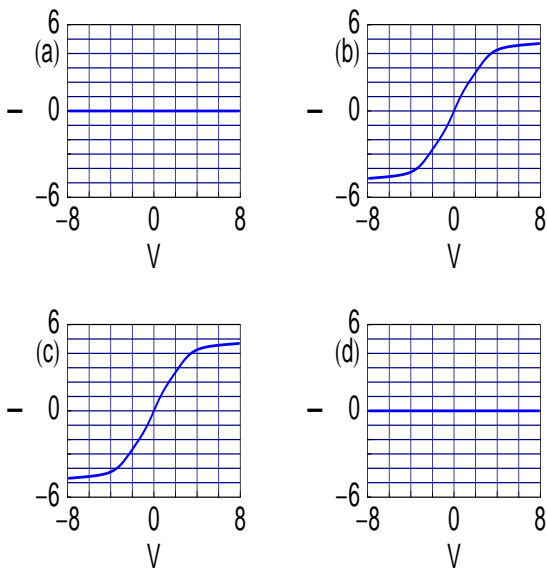}}\par}
\caption{(Color online). $I$-$V$ curves for the interferometer with
side-coupled quantum dots. (a) $V_a=V_b=0$, (b) $V_a=2$ and $V_b=0$,
(c) $V_a=0$ and $V_b=2$ and (d) $V_a=V_b=2$. Other parameters are,
$\epsilon_1=\epsilon_2=\epsilon_3=\epsilon_4=0$, $t=3$, $t_a=t_b=3$,
$\tau_S=\tau_D=2.5$, $\phi=0.5$, $\epsilon^{\prime}=0$ and $t^{\prime}=4$.}
\label{xorcurr}
\end{figure}
flux $\phi$ at $\phi_0/2$ i.e., $0.5$ in our chosen unit $c=e=h=1$.
As representative examples, in Fig.~\ref{xorcond} we show the variation
of the conductance $g$ as a function of the injecting electron energy
$E$, where (a), (b), (c) and (d) represent the results for the different
cases of the gate voltages $V_a$ and $V_b$ applied in the two side-attached
quantum dots, respectively. When both the two inputs $V_a$ and $V_b$ are
identical to zero, i.e., both the two inputs are low, the conductance
becomes exactly zero for the entire energy range (see Fig.~\ref{xorcond}(a)).
This reveals that the electron conduction through the interferometer is 
not possible for this particular case. Similar behavior is also observed 
for the typical case when both the two inputs are high i.e., $V_a=V_b=2$.
In this situation also the electron conduction from the source to drain
through the interferometer is not possible for the whole energy range
(see Fig.~\ref{xorcond}(d)). On the other hand, for the rest two cases
i.e., when any one of the two inputs is high and other one is low i.e.,
either $V_a=2$ and $V_b=0$ (Fig.~\ref{xorcond}(b)) or $V_a=0$ and $V_b=2$
(Fig.~\ref{xorcond}(c)), the conductance exhibits resonances for some 
particular energies. Thus for both these two cases the electron conduction 
takes place across the interferometer. Now we justify the dependences of 
the gate voltages on the electron transport for these four different cases. 
The probability amplitude of getting an electron across the interferometer 
depends on the quantum interference of the electronic waves passing through 
the upper and lower arms of the interferometer. For the symmetrically 
connected interferometer i.e., when the two arms of the interferometer 
are identical with each other, the probability amplitude is exactly
zero ($T=0$) for the flux $\phi=\phi_0/2$. This is due to the result of
\begin{table}[ht]
\begin{center}
\caption{XOR gate response of the quantum interferometer with side-coupled
dots. The current $I$ is computed at the bias voltage $6.02$.}
\label{table1}
~\\
\begin{tabular}{|c|c|c|}
\hline \hline
Input-I ($V_a$) & Input-II ($V_b$) & Current ($I$) \\ \hline
$0$ & $0$ & $0$ \\ \hline
$2$ & $0$ & $4.568$ \\ \hline
$0$ & $2$ & $4.568$ \\ \hline
$2$ & $2$ & $0$ \\ \hline \hline
\end{tabular}
\end{center}
\end{table}
the quantum interference among the two waves in the two arms of the 
interferometer, which can be obtained through the few simple mathematical 
steps. Thus for the cases when both the two inputs ($V_a$ and $V_b$) 
are either low or high, the transmission probability drops to zero. 
While, for the two other cases, the symmetry of the two arms of the 
interferometer is broken by applying the gate voltage either in the 
dot $a$ or in $b$, and therefore, the non-zero value of the transmission 
probability is achieved which reveals the electron conduction across 
the interferometer. Thus we can predict that the electron conduction 
takes place across the interferometer if one, and only one, of the two
inputs to the gate is high, while if both the inputs are low or both 
are high the conduction is no longer possible. This feature clearly 
demonstrates the XOR gate behavior.

The XOR gate response can be much more clearly noticed by studying the 
$I$-$V$ characteristics. The current $I$ passing through the interferometer
is computed from the integration procedure of the transmission function
$T$ as prescribed in Eq.~(\ref{equ8}). The transmission function varies 
exactly similar to that of the conductance spectrum, differ only in 
magnitude by the factor $2$ since the relation $g=2T$ holds from the 
Landauer conductance formula Eq.~(\ref{equ1}). As representative examples,
in Fig.~\ref{xorcurr} we display the variation of the current $I$ as a 
function of the applied bias voltage $V$ for the four different cases
of the two gate voltages $V_a$ and $V_b$. In the particular cases when 
both the two inputs are identical to each other, either low 
(Fig.~\ref{xorcurr}(a)) or high (Fig.~\ref{xorcurr}(d)), the current 
is zero for the complete range of the bias voltage $V$. This behavior 
is clearly understood from the conductance spectra, Figs.~\ref{xorcond}(a) 
and (d), since the current is computed from the integration method of 
the transmission function $T$. For the other two cases when only one 
of the two inputs is high and other is low, a high output current is 
obtained which are clearly described in Figs.~\ref{xorcurr}(b) and (c). 
From these $I$-$V$ curves the behavior of the XOR gate response is 
nicely observed. To make it much clear, in Table~\ref{table1}, we 
present a quantitative estimate of the typical current amplitude, 
computed at the bias voltage $V=6.02$. It shows that, $I=4.568$ only 
when any one of the two inputs is high and other is low, while for 
the other cases when either $V_a=V_b=0$ or $V_a=V_b=2$, the current 
achieves the value $0$.

\section{Concluding remarks}

To summarize, we have studied electron transport in a quantum interferometer
with side-coupled quantum dots. The interferometer, threaded by a magnetic
flux $\phi$, is attached symmetrically to two semi-infinite $1$D metallic
electrodes and two gate voltages, viz, $V_a$ and $V_b$, are applied,
respectively, in the two dots those are treated as the two inputs of the 
XOR gate. The system is described by the tight-binding Hamiltonian 
and all the calculations are done in the Green's function formalism. We 
have numerically computed the conductance-energy and current-voltage 
characteristics as functions of the interferometer-to-dots coupling 
strengths, magnetic flux and gate voltages. We have described the essential
features of the electron transport in two parts. In the first part, we
have addressed the existence of the anti-resonant states, those are not
available in the traditional scattering problems of potential barriers.
On the other hand, in the second part, we have explored the XOR gate
response for this particular model. Very interestingly we have noticed 
that, for the half flux-quantum value of $\phi$ ($\phi=\phi_0/2$), a high 
output current ($1$) (in the logical sense) appears if one, and only one, 
of the inputs to the gate is high ($1$). On the other hand, if both the 
two inputs are low ($0$) or both are high ($1$), a low output current ($0$) 
appears. It clearly demonstrates the XOR gate behavior, and, this aspect 
may be utilized in designing a tailor made electronic logic gate. 

The importance of this article is concerned with (i) the simplicity of the
geometry and (ii) the smallness of the size. To the best of our knowledge 
the XOR gate response in such a simple low-dimensional system has not been 
addressed earlier in the literature.

\end{document}